\documentclass[prb,nopacs,twocolumn,preprintnumbers,amsmath,amssymb]{revtex4}

\usepackage{epsfig}
\usepackage{graphicx}
\usepackage{dcolumn}
\usepackage{color}
\usepackage{bm}

\begin{document}

\title{Spin-orbit coupling assisted by flexural phonons in graphene}

\author{H. Ochoa,$^1$ A. H. Castro Neto,$^{2,3}$ V. I. Fal'ko,$^{4,5}$ F. Guinea$^1$}

{\affiliation{$^1$Instituto de Ciencia de Materiales de Madrid. CSIC. Sor Juana In\'es de la Cruz 3. 28049 Madrid. Spain. \\
$^2$Graphene Research Centre and Department of Physics, National University of Singapore, 6 Science Drive 2, 117546, Singapore. \\
$^3$Department of Physics, Boston University, 590 Commonwealth Ave., Boston MA 02215, USA. \\
$^4$Physics Department, Lancaster University, Lancaster, LA1 4YB, UK. \\
$^5$DPMC, University of Geneva, 24 Quai Ernest-Ansermet, CH1211 Geneve 4, Switzerland }

\begin{abstract}
We analyze the couplings between spins and phonons in graphene. We present a complete analysis of the possible couplings between spins and flexural, out of plane, vibrations. From tight-binding models we obtain analytical and numerical estimates of their strength. We show that dynamical effects, induced by quantum and thermal fluctuations, significantly enhance the spin-orbit gap.

\end{abstract}

\maketitle
\section{Introduction}
Graphene combines in an unique way structural and electronic properties not found in other materials\cite{Novoselov_etal,CastroNeto_etal}. Its special electronic structure made graphene the initial model for a two dimensional topological insulator.\citep{Haldane,Kane_Mele,Hasan_Kane} Numerical calculations suggested that the gap which is the hallmark of a topological insulator is in the micro Kelvin range, too small to be experimentally observed.\cite{Huertas-Hernando_etal_prb,YYQZF07,Min_etal,Gmitra_etal} Recently, different approaches based on heavy adatoms deposition have been proposed in order to engineer a topological state.\citep{Weeksetal,Jin_etal} Experimentally, these proposals are difficult to be implemented, since the presence of adatoms implies a $z\rightarrow -z$ (mirror) symmetry breaking which induces a Rashba-like spin-orbit coupling (SOC).\citep{CastroNeto_Paco} These calculations are also the main guide for the interpretation of spin transport experiments. The calculations assumed lattice made up of ions of infinite mass, and ignored the lattice degrees of freedom.

In the present work, we analyze the coupling of spins to lattice vibrations in graphene, with emphasis on flexural modes, which are unique to a two dimensional membrane, and have lower frequencies than conventional in-plane phonons. Flexural phonons have a great impact on the SOC since out-of-plane distortions of the lattice hybridize $\pi$ orbitals with higher orbitals of carbon, leading to a first order contribution in the spin-orbit interaction strength, contrary to in-plane distortions, whose contribution is at least quadratic, as it happens with the intrinsic SOC.\cite{Huertas-Hernando_etal_prb,Min_etal} The main results of our paper are: i) A complete characterization of all possible couplings to flexural modes, including analytical expressions for each of them, and ii) An analysis of the effect of dynamic flexural modes on the Kane-Mele spin-orbit coupling, which defines the range of parameters where the topological insulator features of graphene can be observed. This analysis is relevant in order to study the effect of ripples in spin transport.\citep{Fratini_etal} Our analysis is also applicable to carbon nanotubes,\citep{Ando} an issue that has become more interesting since carbon nanotubes have been recently proposed as a possible platform for hosting Majorana fermions.\citep{Sau_Tewari,Klinovaja_etal}

We first present a symmetry analysis of the possible couplings between spins and lattice vibrations. This study can also be applied to static out of plane deformations. Next, in Section III, we make numerical estimates of the couplings, based on simple tight binding models. Section IV discusses the change induced in the spin-orbit couplings due to the presence of lattice excitations, due to quantum and thermal fluctuation. Finally, Section V contains a summary of the more relevant results. A number of mathematical details are included in the Appendices.

\section{Symmetry analysis and couplings}
 
The low energy sector of the electronic spectrum lies on the two inequivalent corners of the Brillouin zone $\mathbf{K}_{\pm}$, known as valleys or Dirac points. In our convention $\mathbf{K}_{\pm}=\pm\left(4\pi/(3\sqrt{3}a),0\right)$, where $a$ is the carbon-carbon distance, see Fig. \ref{fig:lattice}. The electronic Hamiltonian reads:
\begin{align}
\mathcal{H}=-iv_F\vec{\Sigma}\cdot\vec{\partial}+\Delta_I\Sigma_z\otimes s_z+\Delta_R\left(\Sigma_x\otimes s_y-\Sigma_y\otimes s_x\right)
\end{align}where $\vec{\Sigma}=\left(\Sigma_x,\Sigma_y\right)$ and $\Sigma_z$ are $4\times4$ matrices associated to the sublattice degree of freedom. This Hamiltonian operate in a space of 8-component Bloch functions $\Psi=(\psi_{A,\mathbf{K}_+,\uparrow},\psi_{B,\mathbf{K}_+,\uparrow},\psi_{B,\mathbf{K}_-,\uparrow},-\psi_{A,\mathbf{K}_-,\uparrow},
\psi_{A,\mathbf{K}_+,\downarrow},\psi_{B,\mathbf{K}_+,\downarrow},\\
\psi_{B,\mathbf{K}_-,\downarrow},-\psi_{A,\mathbf{K}_-,\downarrow})^T$. The first term corresponds to the massless Dirac Hamiltonian, the second term is the Kane-Mele mass which arises due to the SO interaction, and the last one is a Rashba-like coupling which is present in the case of a mirror symmetry breaking. Both $\Delta_I$ and $\Delta_R$ are weak, $1-15$ $\mu$eV in the former case according to previous estimates.\citep{Huertas-Hernando_etal_prb,Min_etal,Gmitra_etal}


\begin{center}
\begin{table}
\begin{tabular}{|c|c|c|}
\hline
Irrep&$z\rightarrow-z$ symmetric&$z\rightarrow-z$ asymmetric\\
\hline
$A_1$&$\Sigma_z\otimes s_z$ &$\Sigma_x\otimes s_y-\Sigma_y\otimes s_x$\\
\hline
$A_2$& &$\Sigma_x\otimes s_x+\Sigma_y\otimes s_y$\\
\hline
$B_2$&$\Lambda_z\otimes s_z$&\\
\hline
$E_1$&$\left(\begin{array}{c}
-\Sigma_y\otimes s_z \\
\Sigma_x\otimes s_z \end{array}\right)$& $\left(\begin{array}{c}
-\Sigma_z\otimes s_y \\
\Sigma_z\otimes s_x \end{array}\right)$\\
\hline
$E_2$& &$\left(\begin{array}{c}
\Sigma_x\otimes s_y+\Sigma_y\otimes s_x \\
\Sigma_x\otimes s_x-\Sigma_y\otimes s_y \end{array}\right)$, $\left(\begin{array}{c}
-\Lambda_z\otimes s_y \\
\Lambda_z\otimes s_x \end{array}\right)$\\
\hline
$E_1'$&$\left(\begin{array}{c}
\Lambda_x\otimes s_z \\
\Lambda_y\otimes s_z \end{array}\right)$&\\
\hline
$G'$& & $\left(\begin{array}{c}
\Lambda_x\otimes s_x\\
\Lambda_x\otimes s_y \\
-\Lambda_y\otimes s_y \\
-\Lambda_y\otimes s_x\end{array}\right)$\\
\hline
\end{tabular}
\caption{Classification of the possible SO coupling terms according to how they transform under the symmetry operations of $C_{6v}''$. Note that all these operators are even under time reversal operation $t\rightarrow -t$.}
\label{tab:so_terms}
\end{table}
\end{center}

We analyze first the SO-coupling assisted electron interaction with flexural phonons from a symmetry group theory approach.\citep{Falko} Before we present the exhaustive analysis, we disclose the results in the following lines. The spin-phonon coupling Hamiltonian reads:\begin{align}
\mathcal{H}_{s-ph}=\mathcal{H}_{A_1}+\mathcal{H}_{B_2}+\mathcal{H}_{G'}
\end{align}
where:
\begin{align}
\mathcal{H}_{A_1}=g_{1}\left(\Sigma_x\otimes s_y-\Sigma_y\otimes s_x\right)\partial_i\partial^i u_{A_1}+\nonumber\\+g_{2}[-\Lambda_z\otimes s_y \left(\partial_x^2 u_{A_1}-\partial_y^2 u_{A_1}\right)+2\Lambda_z\otimes s_x\partial_x\partial_yu_{A_1}]+\nonumber\\+g_{3}[\left(\Sigma_x\otimes s_y+\Sigma_y\otimes s_x\right) \left(\partial_x^2 u_{A_1}-\partial_y^2 u_{A_1}\right)+\nonumber\\+2\left(\Sigma_x\otimes s_x-\Sigma_y\otimes s_y\right)\partial_x\partial_yu_{A_1}]
\label{eq:e-A1}
\end{align}
\begin{align}
\mathcal{H}_{B_2}=g_{4}\Sigma_z\otimes s_z (u_{B_2})^2
\label{eq:e-B2}
\end{align}
\begin{align}
\mathcal{H}_{G'}=g_{5}[(\Lambda_x\otimes s_x) u_1+(\Lambda_x\otimes s_y) u_2-\nonumber\\-(\Lambda_y\otimes s_y) u_3-(\Lambda_y\otimes s_x) u_4]
\label{eq:e-G}
\end{align}The main ingredients in order to construct these couplings, electronic operators and symmetry-adapted phonon fields, are summarized in Tab. \ref{tab:so_terms} and Tab. \ref{tab:phonons} respectively. Next, we construct these couplings step by step.

\begin{center}
\begin{table}
\begin{tabular}{|c|c|}
\hline
Irrep&Z phonon mode ($z\rightarrow-z$ asymmetric)\\
\hline
&\\
$A_1$&\includegraphics[width=0.25\columnwidth]{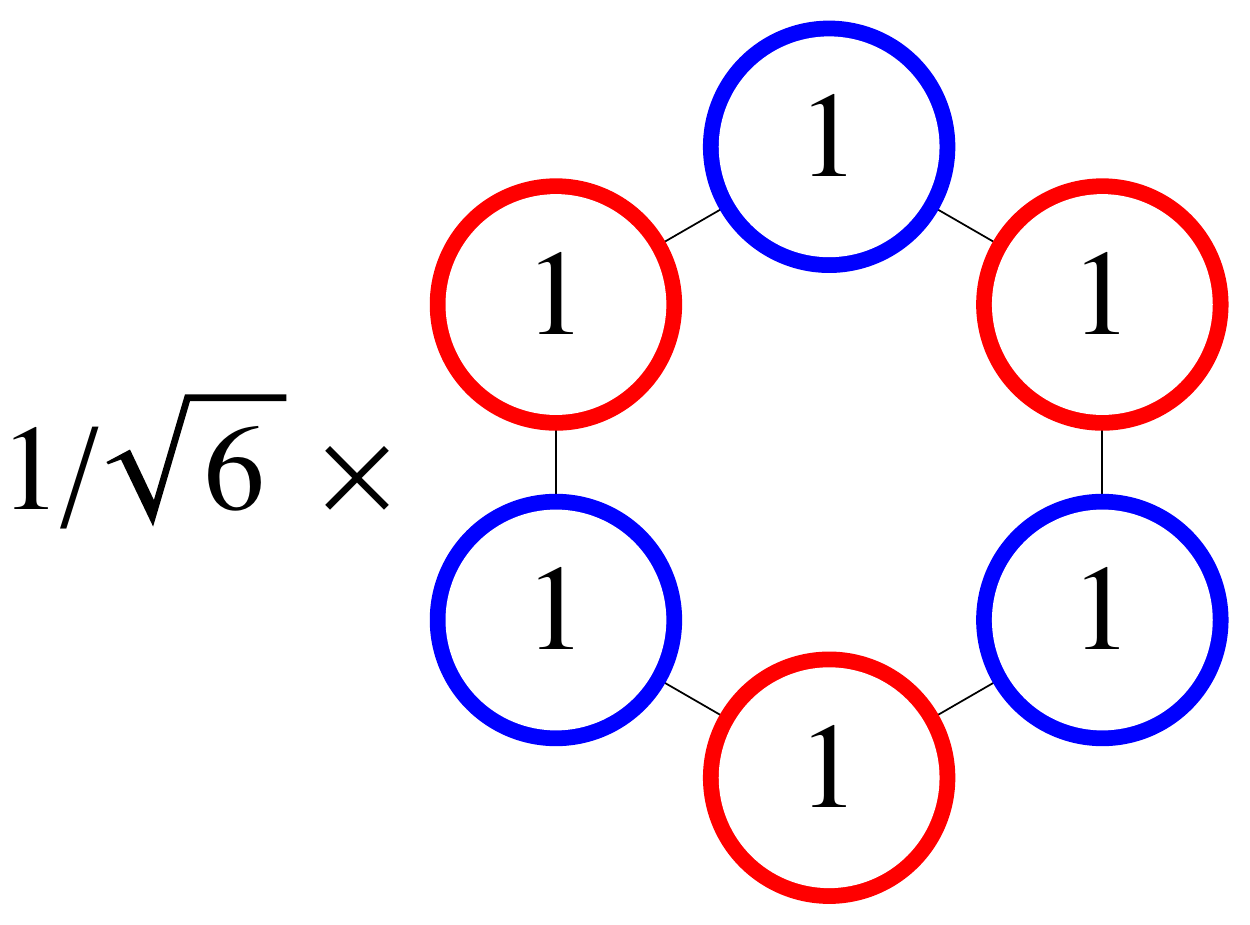}\\
&\\
\hline
&\\
$B_2$&\includegraphics[width=0.25\columnwidth]{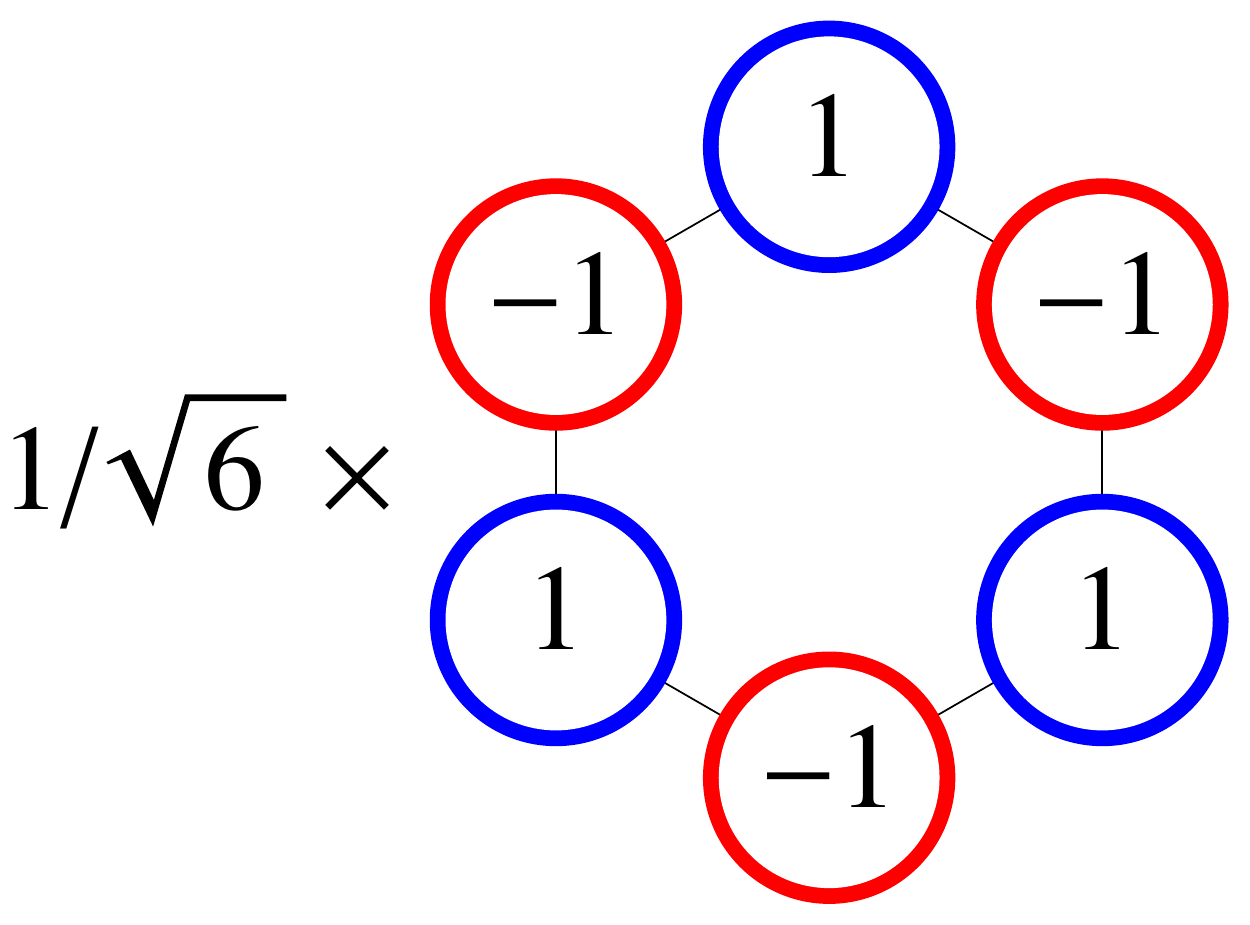}\\
&\\
\hline
&\\
$G'$&\includegraphics[width=0.25\columnwidth]{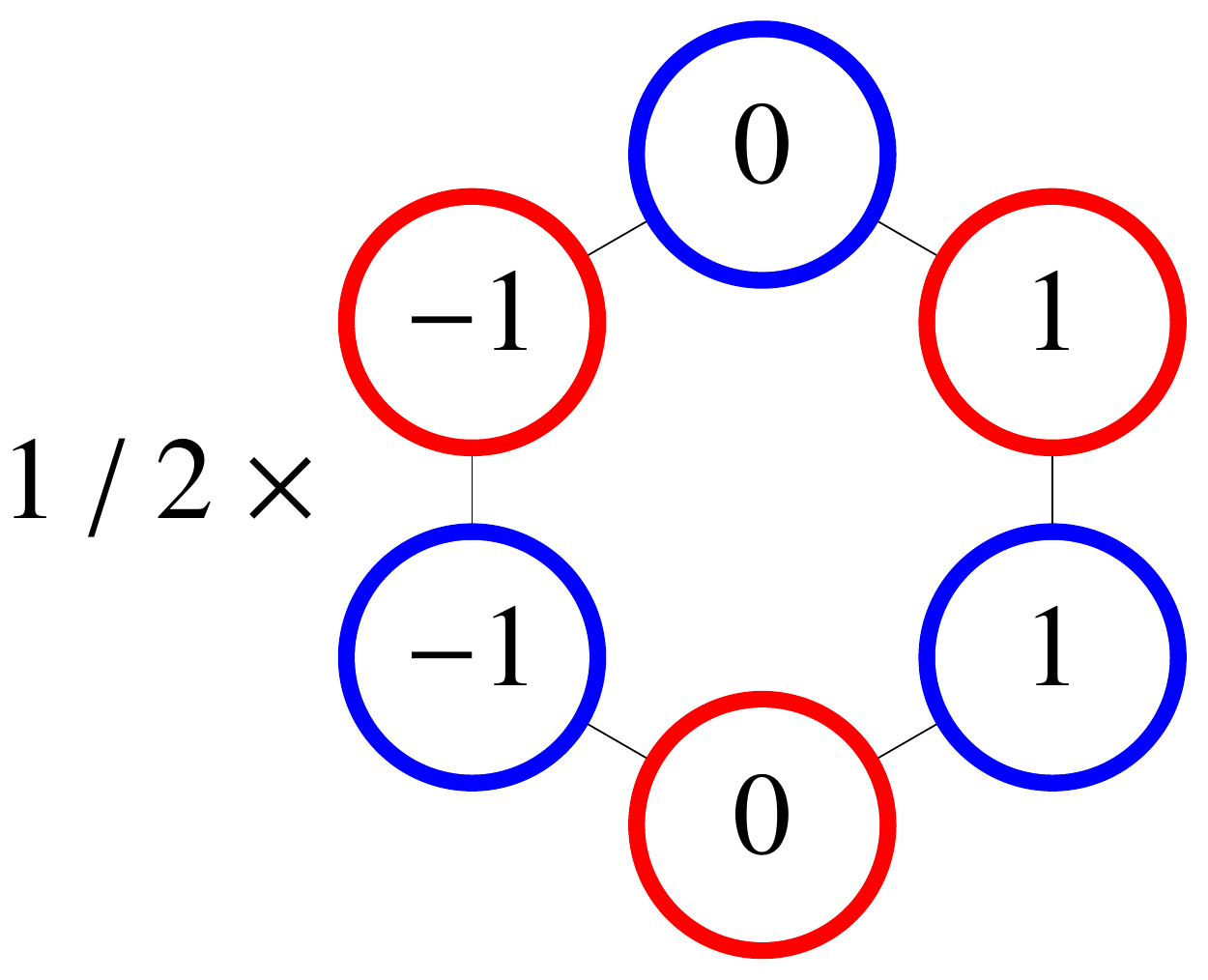}\\ &\includegraphics[width=0.28\columnwidth]{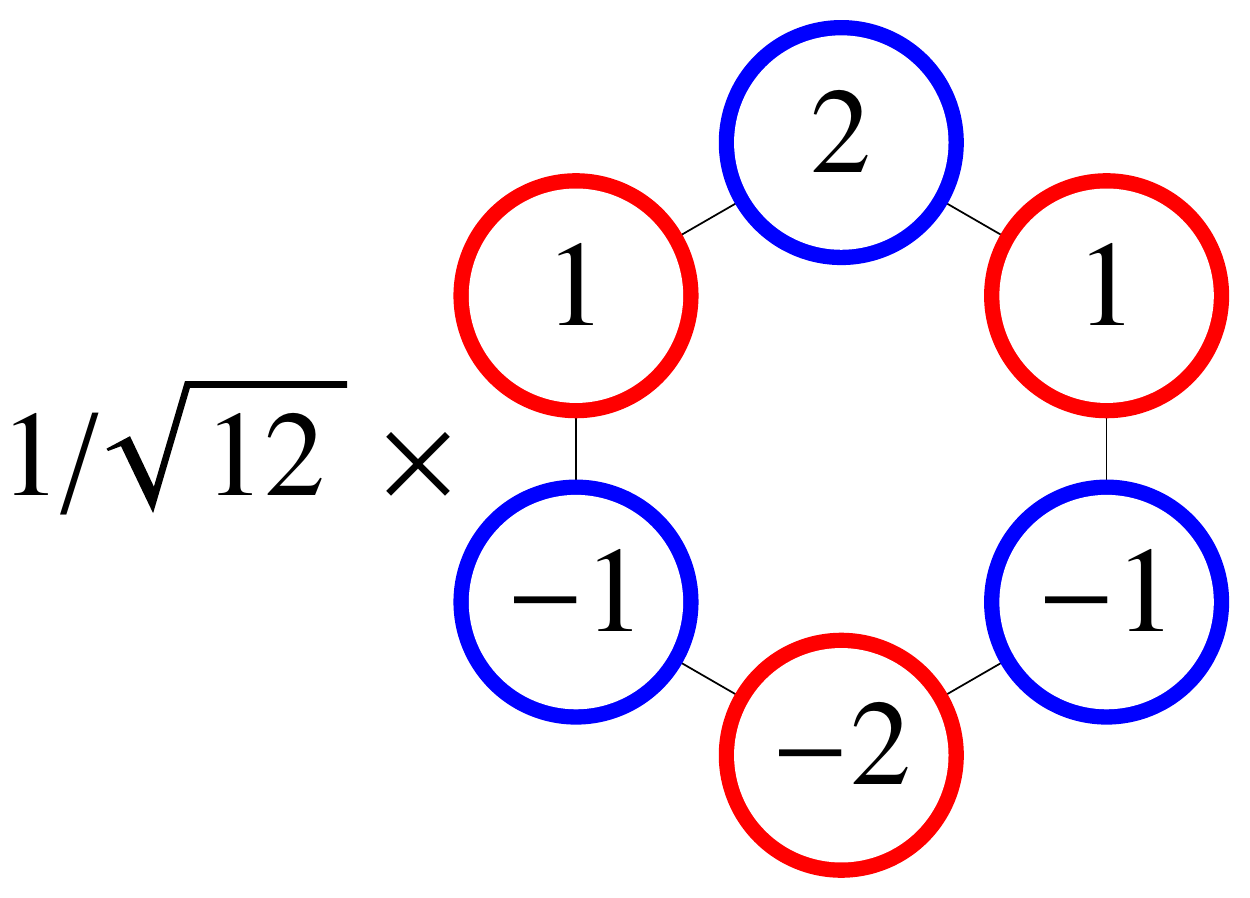}\\
&\includegraphics[width=0.25\columnwidth]{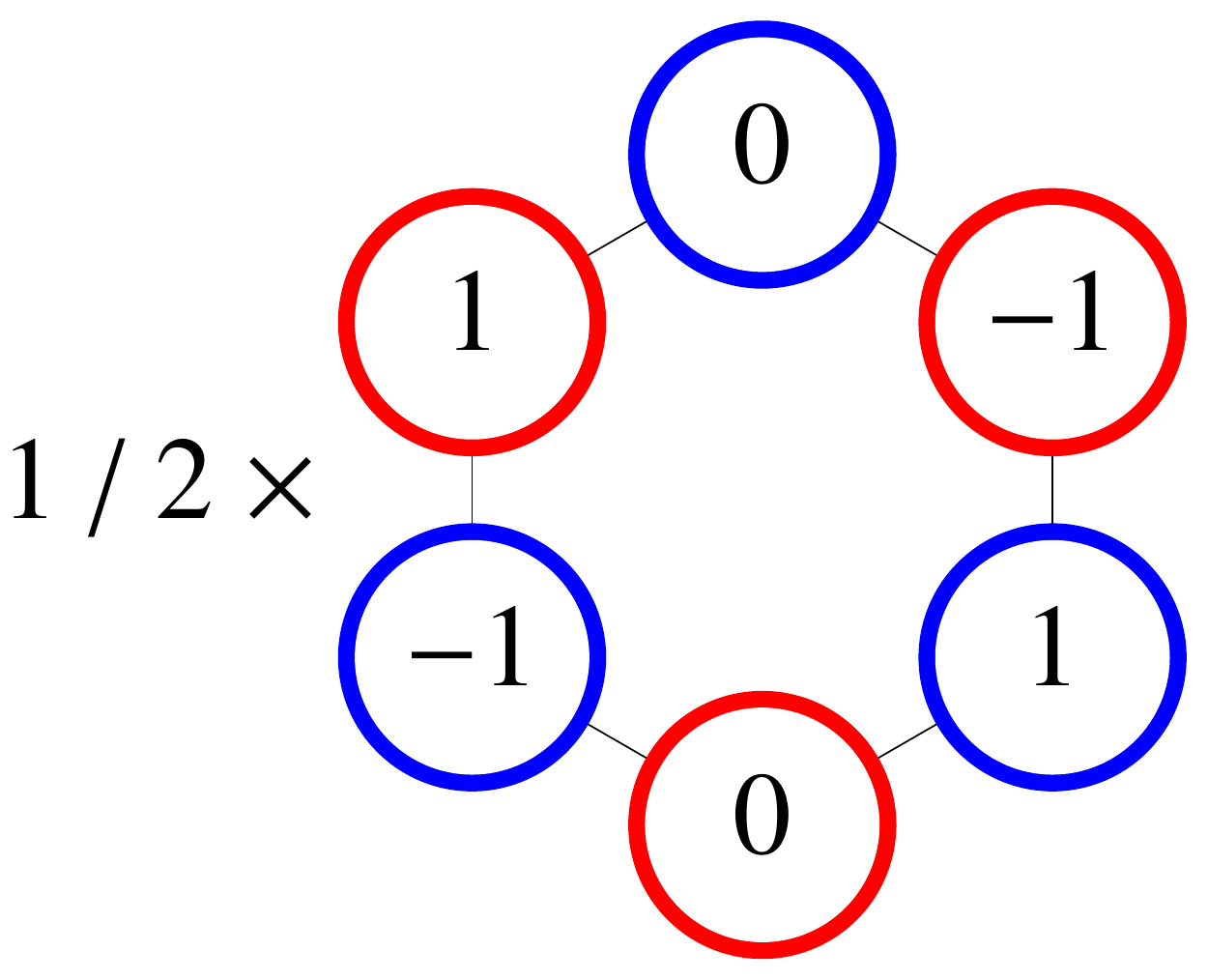}\\
&\includegraphics[width=0.28\columnwidth]{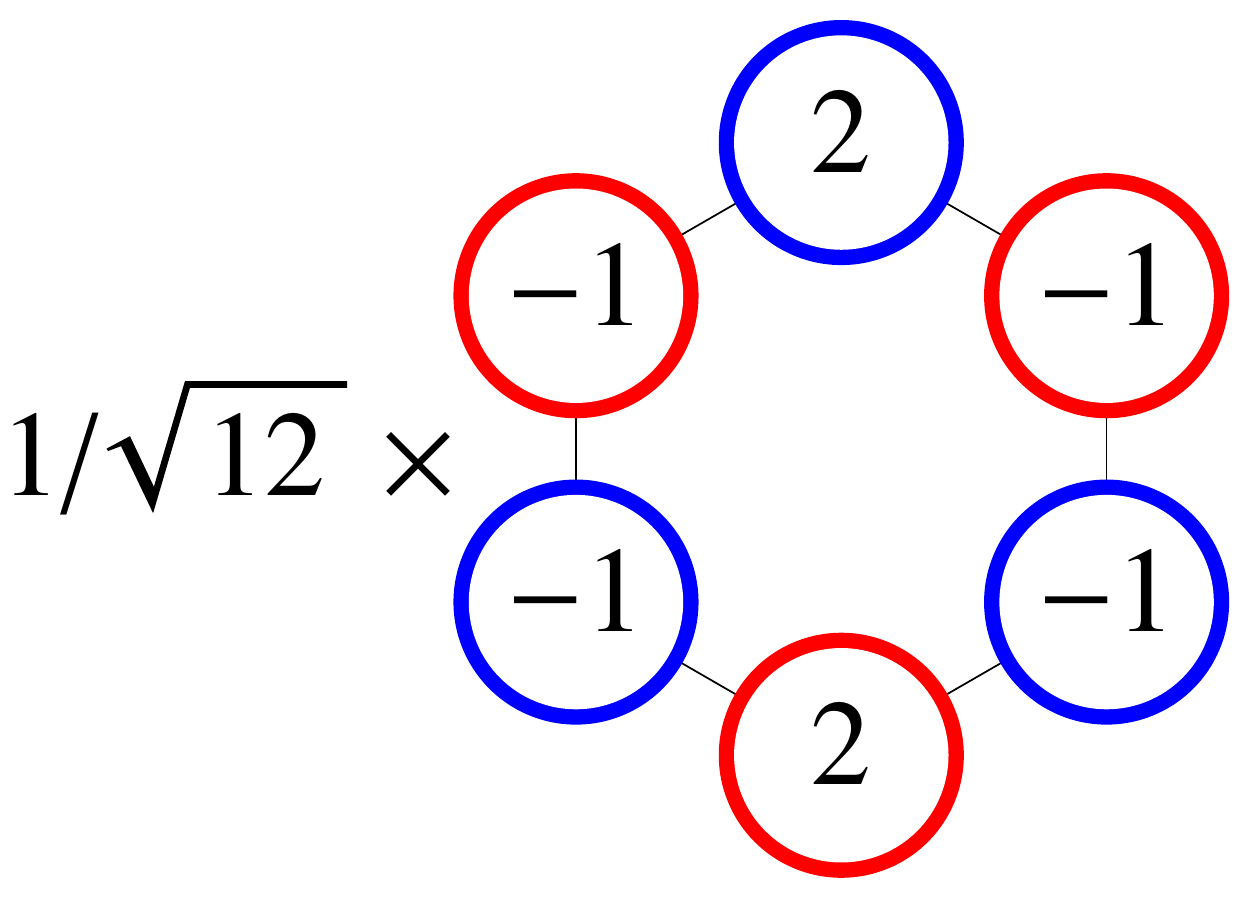}\\
&\\
\hline
\end{tabular}
\caption{Classification of flexural (Z) phonon modes according to how they transform under the symmetry operations of $C_{6v}''$. The numbers indicate the out-of-plane displacement of the atom within the 6-atoms unit cell.}
\label{tab:phonons}
\end{table}
\end{center}

\subsection*{Electronic operators}

The point group of the graphene crystal is $C_{6v}$, which contains 12 elements: the identity, five rotations and six reflections in planes perpendicular to the crystal plane. Instead of dealing with degenerate states at two inequivalent points one can enlarge the unit cell in order to contain six atoms, in such a way that $\mathbf{K}_{\pm}$ are mapped onto the $\mathbf{\Gamma}$ point (see Fig. \ref{fig:lattice} b)). From the point of view of the lattice symmetries, this means that the two elementary translations ($t_{\mathbf{a}_1}$, $t_{\mathbf{a}_2}$) are factorized out of the translation group and added to the point group $C_{6v}$, which becomes $C_{6v}''=C_{6v}+t_{\mathbf{a}_1}\times C_{6v}+t_{\mathbf{a}_2}\times C_{6v}$. If we do not consider the spin degree of freedom, the $\pi$ electronic states at $\mathbf{K}_{\pm}$ transforms according to the 4-dimensional $G'$ irreducible representation of $C_{6v}''$. Then, in order to write down the low-energy electronic Hamiltonian we have to consider the 16 Hermitian operators acting in a 4-dimensional space. We define two different sets of $4\times4$ hermitian matrices $\left\{\Sigma_i\right\}$, $\left\{\Lambda_i\right\}$ associated to sublattice and valley degrees of freedom respectively, so the set $\left\{\mathcal{I},\Sigma_i,\Lambda_i,\Sigma_i\cdot\Lambda_j\right\}$ provides a representation of the algebra of generators of $U(4)$. In the basis introduced before $\Sigma_i$ and $\Lambda_i$ matrices are odd under the time reversal operation $t\rightarrow-t$, see Appendix A. If we consider the spin degree of freedom then the $\pi$ electronic states at $\mathbf{K}_{\pm}$ transform according to the 8-dimensional representation $G'\times D_{1/2}$, where $D_{1/2}$ is the spinor representation associated to the spinorial part of the wave function. Thus, we introduce a set of Pauli matrices $\left\{s_i\right\}$ associated to the spin degree of freedom, also odd under time reversal operation. Importantly, we are now introducing a pseudovector in the 3-dimensional space, meaning that the operators which contain $s_z$ are even under a reflection in the graphene plane (mirror symmetry operation, $z\rightarrow-z$), whereas the operators which contain the in-plane components are odd. The result of this symmetry based approach is summarized in Tab. \ref{tab:so_terms}, where all the possible time reversal symmetric terms which involve spin operators are classified according to how they transforms under the symmetry operations of $C_{6v}''$. These matrices correspond to all the allowed SO coupling terms within the low-energy description.

\begin{figure}
\begin{centering}
\includegraphics[width=0.46\columnwidth]{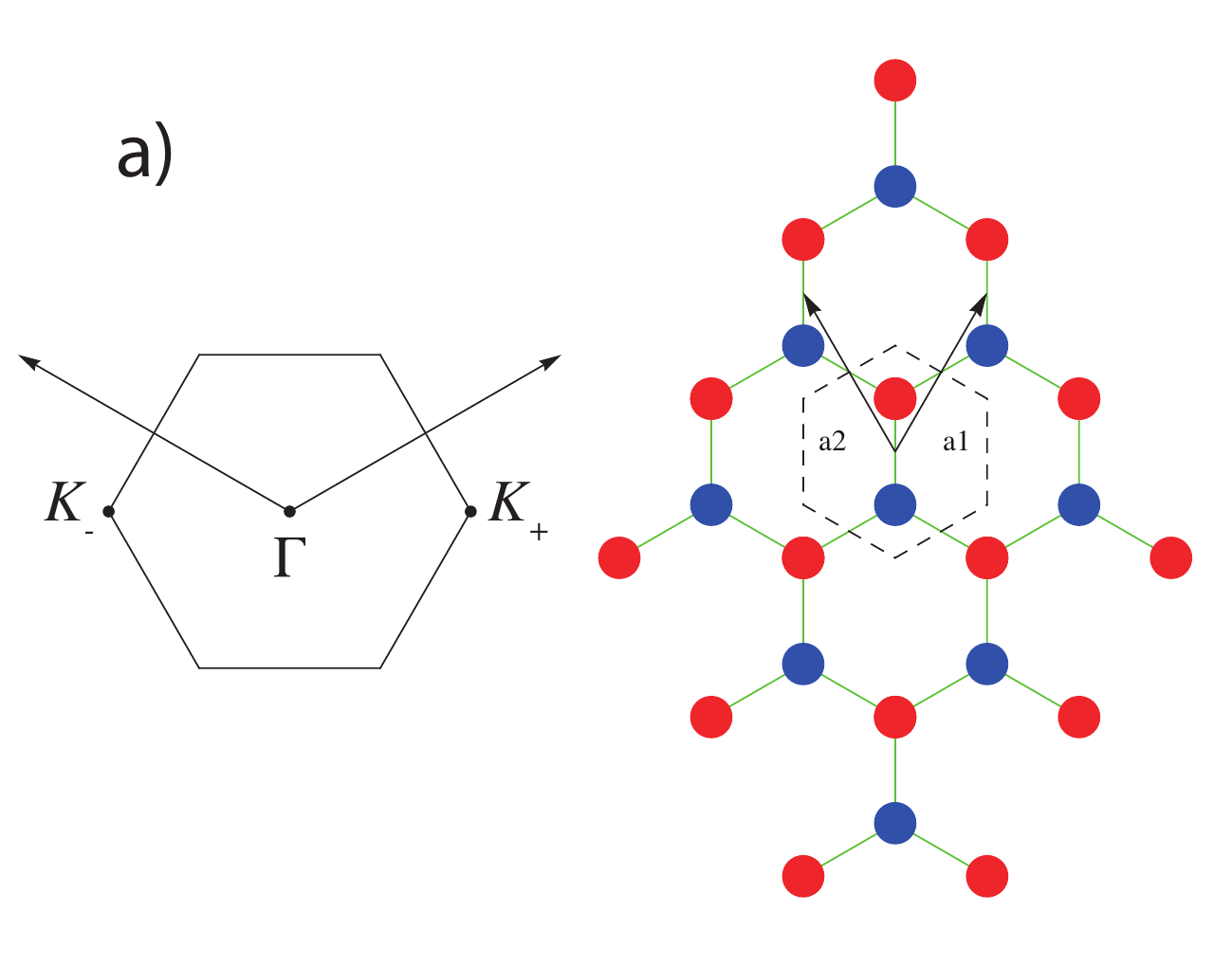}
\includegraphics[width=0.49\columnwidth]{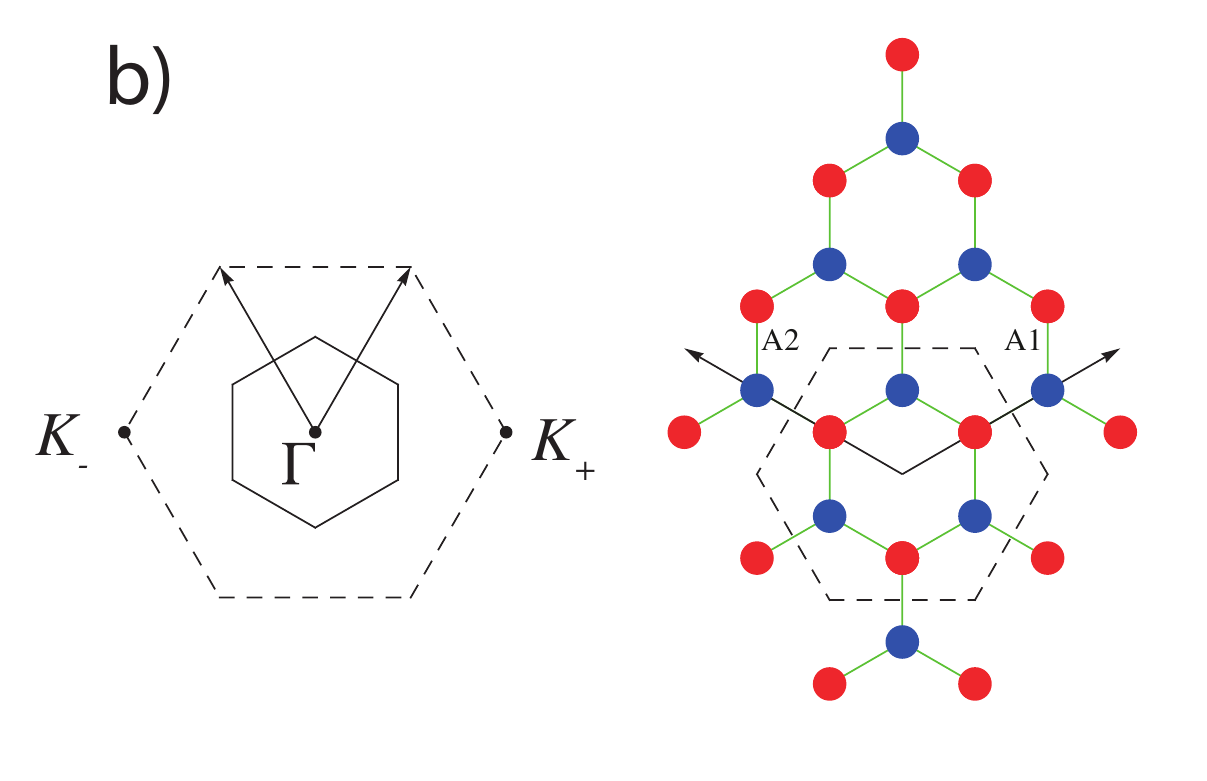}
\par\end{centering}
\caption{\label{fig:lattice}a) Real space lattice and Brillouin zone of flat graphene with 2 atoms per unit cell. b) The same with 6 atoms per unit cell. Note that $\mathbf{K}_{\pm}$ are now equivalent to $\mathbf{\Gamma}$.}
\end{figure}

\subsection*{Phonon modes}

In the lattice with 6 atoms per unit cell, the amplitude of a flexural phonon mode at $\mathbf{\Gamma}$ ($\equiv \mathbf{K}_{\pm}$) is given by a 6-component vector whose entries are associated to the displacements of each sublattice atoms: $\left|h\right\rangle=(h_{A1},h_{B1},h_{A2},h_{B2},h_{A3},h_{B3})$. This vector belongs to a 6-dimensional representation of $C_{6v}''$ which can be reduced as $A_1+B_2+G'$. The polarization vectors associated to the 1-dimensional irreducible representations correspond to the acoustic (ZA) and optical (ZO) modes at the original $\mathbf{\Gamma}$ point (with 2 atoms per unit cell), whose polarization vectors in the 6 atoms basis read:
\begin{align}
\left|A_1\right\rangle=\frac{1}{\sqrt{6}}(1,1,1,1,1,1)\nonumber\\
\left|B_2\right\rangle=\frac{1}{\sqrt{6}}(1,-1,1,-1,1,-1)
\label{eq:Gamma_phonons}
\end{align}
The 4-dimensional irreducible representation corresponds to the 4 degenerate modes at $\mathbf{K}_{\pm}$ points, whose polarization vectors read:
\begin{align}
\left|A,\mathbf{K}_+\right\rangle=\frac{1}{\sqrt{3}}(1,0,e^{i\frac{2\pi}{3}},0,e^{-i\frac{2\pi}{3}},0)\nonumber\\
\left|B,\mathbf{K}_+\right\rangle=\frac{1}{\sqrt{3}}(0,1,0,e^{i\frac{2\pi}{3}},0,e^{-i\frac{2\pi}{3}})\nonumber\\
\left|A,\mathbf{K}_-\right\rangle=\frac{1}{\sqrt{3}}(1,0,e^{-i\frac{2\pi}{3}},0,e^{i\frac{2\pi}{3}},0)\nonumber\\
\left|B,\mathbf{K}_-\right\rangle=\frac{1}{\sqrt{3}}(0,1,0,e^{-i\frac{2\pi}{3}},0,e^{i\frac{2\pi}{3}})
\label{eq:G_phonons}
\end{align}
Note that $\left|A/B,\mathbf{K}_-\right\rangle=\left(\left|A/B,\mathbf{K}_+\right\rangle\right)^*$. However, we must consider the real linear combinations of the vectors of Eq. \eqref{eq:G_phonons} which transforms according to $G'$ in order to construct the couplings. These are:
\begin{align}
\left|1\right\rangle=\frac{i}{2}\left[-\left|A\mathbf{K}_+\right\rangle+\left|A\mathbf{K}_-\right\rangle-\left|B\mathbf{K}_+\right\rangle+\left|B\mathbf{K}_-\right\rangle\right]\nonumber\\
\left|2\right\rangle=\frac{1}{2}\left[\left|A\mathbf{K}_+\right\rangle+\left|A\mathbf{K}_-\right\rangle-\left|B\mathbf{K}_+\right\rangle-\left|B\mathbf{K}_-\right\rangle\right]\nonumber\\
\left|3\right\rangle=\frac{i}{2}\left[-\left|A\mathbf{K}_+\right\rangle+\left|A\mathbf{K}_-\right\rangle+\left|B\mathbf{K}_+\right\rangle-\left|B\mathbf{K}_-\right\rangle\right]\nonumber\\
\left|4\right\rangle=\frac{1}{2}\left[\left|A\mathbf{K}_+\right\rangle+\left|A\mathbf{K}_-\right\rangle+\left|B\mathbf{K}_+\right\rangle+\left|B\mathbf{K}_-\right\rangle\right]
\label{eq:G_phonons_real}
\end{align}
The polarization vectors of Eq. \eqref{eq:Gamma_phonons}, together with he vectors of Eq. \eqref{eq:G_phonons_real}, form a symmetry adapted basis, in such a way that the displacement vector of a flexural mode can be written as $\left|h\right\rangle=u_{A_1}\left|A_1\right\rangle+u_{B_2}\left|B_2\right\rangle+u_1\left|1\right\rangle+u_2\left|2\right\rangle+u_3\left|3\right\rangle+u_4\left|4\right\rangle$, where $u_i$ are the symmetry adapted (real) displacement fields. The results of this analysis are summarized in Tab. \ref{tab:phonons}.

\subsection*{Spin-phonon couplings}

This analysis allows us to identify the SO assisted electron coupling with flexural phonons at the center and the corners of the Brillouin zone. The spin-phonon interaction Hamiltonian can be expanded in powers of the phonon displacement fields and their derivatives, in such a way that the displacement fields (and the derivatives) are paired with the electronic operators corresponding to the same irreducible representation, and taking into account that these combinations must be even under the operation $z\rightarrow-z$.  The couplings of Eqs. \eqref{eq:e-A1}-\eqref{eq:e-G} correspond to the leading terms in such expansion.

Since a uniform translation of the crystal cannot affect the electron motion, it is clear that it can couple to acoustic phonons (at $\mathbf{\Gamma}$, $A_1$ phonons) only through spatial derivatives of the corresponding displacement field. Moreover, if we consider the graphene as a continuum surface, only out-of-plane distortions which generate extrinsic curvature can couple to electron spin through the SO interaction. This implies that the leading term must depend on second derivatives $\partial_i\partial_j u_{A_1}$. Since $\partial_i\partial^i u_{A_1}$ transforms according to $A_1$, and $\left(\partial_x^2u_{A_1}-\partial_y^2u_{A_1},2\partial_x\partial_yu_{A_1}\right)$ forms a doublet which transforms according to $E_2$, in principle three different couplings are allowed by the symmetries. We obtain Eq. \ref{eq:e-A1}.

In the case of optical phonons at the center of the Brillouin zone, note that there is no term which transforms according to the $B_2$ irreducible representation in the column of $z\rightarrow -z$ asymmetric operators, which means that the leading term must be quadratic on the phonon displacement fields. Since $B_2\times B_2=A_1$ we obtain a Kane-Mele-like coupling term, Eq. \ref{eq:e-B2}.

Finally, in the case of flexural phonons at the corners of the Brillouin zone, where both acoustic and optical branches are degenerate, the coupling reads as Eq. \ref{eq:e-G}.

\section{Tight-binding model}
The strength of these couplings can be estimated from a tight-binding model. We assume the convention of Fig \ref{fig:lattice}. We choose the simplest tight-binding model with 4 orbitals $\left\{s,p_x,p_y,p_z\right\}$ per carbon atom and only nearest neighbors hoppings. We neglect the effect of $d$ orbitals\citep{Gmitra_etal} because all these couplings are first order in the SO interaction constant, as we are going to see. The Hamiltonian can be written as:
\begin{align}
\mathcal{H}_{TB}=\sum_{i,\lambda}t_{\lambda}c_{i,\lambda}^{\dagger}c_{i,\lambda}+\left[\sum_{<i,j>}\sum_{\lambda,\lambda'}t_{\lambda,\lambda'}^{ij}c_{i,\lambda}^{\dagger}c_{j,\lambda'}+H.C.\right]
\end{align}
where the latin indices label the sites of the carbon atoms and $\lambda=s,p_x,p_y,p_z$ labels the orbitals considered in the calculation. For the on-site energies we take $t_s=\epsilon_s$, $t_{p_i}=\epsilon_p$. The two-center matrix elements can be computed within the Slater-Koster approximation as it is indicated in Tab. \ref{tab:matrix_elements}. We take (in eV):\citep{Tomanek_Louie,Tomanek_Schluter} $\epsilon_s=-7.3$, $\epsilon_p=0$, $V_{ss\sigma}=-3.63$, $V_{sp\sigma}=4.2$, and $V_{pp\sigma}=5.38$, and $V_{pp\pi}=-2.24$. We describe the spin-obit interaction in terms of the Hamiltonian $\mathcal{H}_{SO}=\Delta \vec{L}\cdot\vec{s}$, where $\vec{L}$ and $\vec{s}$ correspond to the orbital angular momentum and spin operators respectively. We take $\Delta=20$ meV.\citep{Serrano_etal}

\begin{center}
\begin{table}
\begin{tabular}{|c|c|}
\hline
&\\
$t_{s,s}^{AB}$&$V_{ss\sigma}$\\
&\\
\hline
&\\
$t_{s,p_i}^{AB}$&$\left(\hat{p}_i\cdot\vec{\delta}\right)V_{sp\sigma}$\\
&\\
\hline
&\\
$t_{p_i,p_j}^{AB}$&$\left(\hat{p}_i\cdot\vec{\delta}\right)\left(\hat{p}_j\cdot\vec{\delta}\right)V_{pp\sigma}+$\\
&$+\left[\left(\hat{p}_i\cdot\hat{p}_j\right)-\left(\hat{p}_i\cdot\vec{\delta}\right)\left(\hat{p}_j\cdot\vec{\delta}\right)\right]V_{pp\pi}$\\
\hline
\end{tabular}
\caption{Two-center matrix elements in the Slater-Koster approximation as function of the tight-binding parameters (see the text). $\hat{p}_i$ represents a unitary vector in the direction of maximum amplitude of the orbital $p_i$, and $\vec{\delta}$ is the vector which connects neighboring sites A and B.}
\label{tab:matrix_elements}
\end{table}
\end{center}

Our aim is to estimate the strength of the effective couplings within the low energy sector of the electronic spectrum. We can define the $\pi$ ($p_z$, low energy sector) and $\sigma$ ($s$, $p_x$, $p_y$, high energy sector) orbital subspaces at least locally (see the discussion next and Fig. \ref{fig:sheets}). Then, the electronic Hamiltonian can be written in the block form:
\begin{align}
\mathcal{H}=\left(\begin{array}{cc}
\mathcal{H}_{\pi} & \mathcal{H}_{\pi\sigma} \\
\mathcal{H}_{\sigma\pi} & \mathcal{H}_{\sigma}
\end{array}\right)
\end{align}
We project out $\sigma$ orbitals by a Schrieffer-Wolf transformation.\citep{Schrieffer-Wolf} We take the Green function $\mathcal{G}=\left(\epsilon-\mathcal{H}\right)^{-1}$, evaluate the block $\mathcal{G}_{\pi}$ associated to the low-energy sector, and use it in order to identify the low-energy effective Hamiltonian. If we define $\mathcal{G}_{\pi,\sigma}^{(0)}=\left(\epsilon-\mathcal{H}_{\pi,\sigma}\right)^{-1}$, then we can write:\begin{align}
\left(\begin{array}{cc}
\mathcal{G}_{\pi} & \mathcal{G}_{\pi\sigma} \\
\mathcal{G}_{\sigma\pi} & \mathcal{G}_{\sigma}
\end{array}\right)=\left(\begin{array}{cc}
\left(\mathcal{G}_{\pi}^{(0)}\right)^{-1} & \mathcal{H}_{\pi\sigma} \\
\mathcal{H}_{\sigma\pi} & \left(\mathcal{G}_{\sigma}^{(0)}\right)^{-1}
\end{array}\right)^{-1}
\end{align}
We obtain $\mathcal{G}_{\pi}=\left[\left(\mathcal{G}_{\pi}^{(0)}\right)^{-1}+\mathcal{H}_{\pi\sigma}\mathcal{G}_{\sigma}^{(0)}\mathcal{H}_{\sigma\pi}\right]^{-1}$,
so $\epsilon-\mathcal{G}_{\pi}^{-1}=\mathcal{H}_{\pi}+\mathcal{H}_{\pi\sigma}\mathcal{G}_{\sigma}^{(0)}\mathcal{H}_{\sigma\pi}$.
In the low energy sector ($\epsilon\approx 0$) the effective Hamiltonian reads:
\begin{align}
\mathcal{H}_{\pi}^{eff}\approx \mathcal{H}_{\pi}-\mathcal{H}_{\pi\sigma}\mathcal{H}_{\sigma}^{-1}\mathcal{H}_{\sigma\pi}
\label{eq:Heff}
\end{align}

Both the SO interaction and the out-of-plane distortions enter in the $\pi-\sigma$ mixing blocks. In the absence of out-of-plane distortions the only coupling allowed by the symmetries has the structure of a Kane-Mele mass. Our estimation from the tight-binding model is
$\Delta_{I}^{flat}=\epsilon_s \Delta^2/\left(18 V_{sp\sigma}^2\right)$
in agreement with Ref. \onlinecite{Min_etal}. We have $\Delta_{I}^{flat}\approx9$ $\mu$eV.

In the case of $B_2$ phonons we have to add to this analysis the effect of a vertical displacement of one sublattice respect to the other. If the calculation is performed with the 6 atoms unit cell one can identify both the coupling with flexural phonons at $\mathbf{K}_{\pm}$ ($G'$). In the case of $A_1$ phonons, since the coupling depends on the second derivatives of the phonon field, the calculation is not so straightforward.

\subsection*{Phonons at $\mathbf{\Gamma}$}

\begin{figure}
\begin{centering}
\includegraphics[width=0.45\columnwidth]{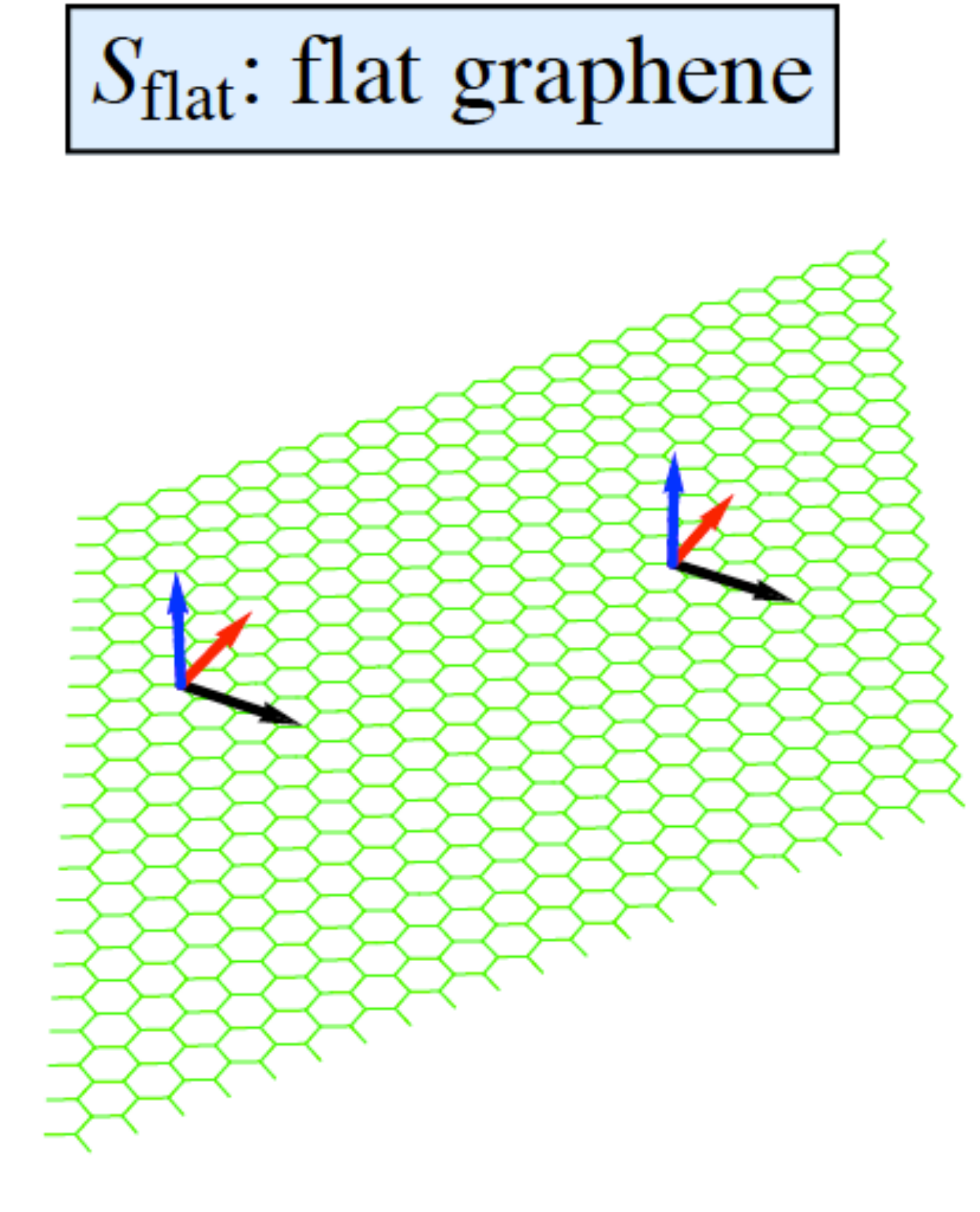}
\includegraphics[width=0.49\columnwidth]{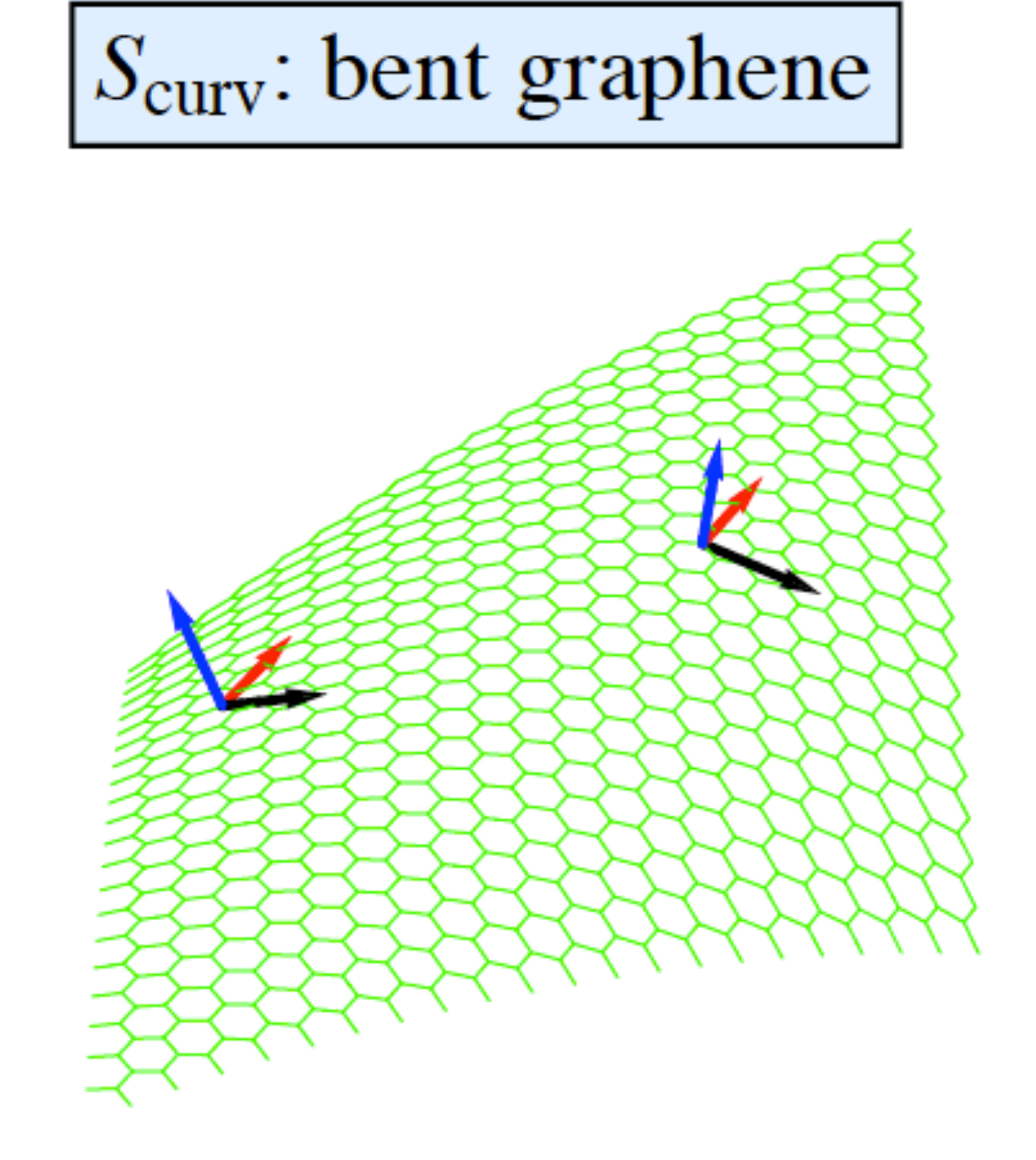}
\par\end{centering}
\caption{\label{fig:sheets}Flat graphene and bent graphene. The arrows represent the $p_x$ (black), $p_y$ (red), and $p_z$ (blue) orbitals. The local basis of atomic $p$ orbitals in the bent graphene is uniquely defined by the isomorphism between them.}
\end{figure}

As we mentioned before, the coupling with acoustic phonons can be inferred from the effect of extrinsic curvature of the graphene sample. The approach that we present here is quite similar to the calculation of the SO coupling in carbon nanotubes. The crucial fact is how to choose the basis of $\pi$ and $\sigma$ orbitals. In the carbon nanotubes calculation the $\sigma$ orbitals are chosen in such a way that they follow the shape of the nanotube, and the $\pi$ orbital is chosen in the radial direction. Here we do essentially the same, but we try to formalize it a little bit. In the low energy limit the graphene sample can be described within a continuum theory. We assume that the position of the carbon atoms lie on a smooth surface, which is valid in the long-wavelength limit. Moreover, we assume that this surface is isometric to a plane. This means that a diffeomorphism $f$ from a flat graphene surface ($S_{flat}$) to a curved graphene surface ($S_{curv}$) exists
in such a way that the metric on the curved surface is pull-backed to the flat one. Thus, the isomorphism $f$ defines an unique way to introduce a local basis for the $p_x$ and $p_y$ atomic orbitals.

Consider graphene in a flat configuration. We introduce unitary vectors $\hat{p}_x$, $\hat{p}_y$ in the direction of maximum amplitude of the orbitals $p_x$, $p_y$ respectively, see Fig. \ref{fig:sheets}. From a geometrical point of view, these vectors are elements of the tangent bundle associated to $S_{flat}$.\citep{Nakahara} At the same time, the axis of maximum localization of the $p_z$ orbital verifies $\hat{p}_z=\hat{p}_x\times\hat{p}_y$. Now consider a curved graphene which is related to the flat graphene by a isomorphism $f$. The push-forward of $f$ maps the tangent bundle of $S_{flat}$ to the tangent bundle
of $S_{curv}$, which means that the vectors $\hat{p}_x$, $\hat{p}_y$ at any position of the bent graphene are uniquely determined by the action of the push-forward of the isomorphism on the original $\hat{p}_x$, $\hat{p}_y$ defined in the flat configuration. And of course, $\hat{p}_z$ in the bent graphene surface is given by the vectorial product of the new $\hat{p}_x$, $\hat{p}_y$. More physically, what we are doing is the following. Consider a sheet of paper. Consider a point on it and write down a small arrow. Now deform the sheet of paper smoothly (folding is not differentiable) without breaking it. The new surface of your paper is then isometric to the original plane. Consider the previous point in the new curved sheet of paper. The vector tangent to the arrow at that point, which is uniquely defined, is actually the push-forwarded original vector on the flat sheet of paper. This way of introducing the local basis in the bent graphene has two advantages: 1) we recover "smoothly" the original basis when we restore the curved graphene to the original flat configuration; 2) we keep the notion of parallelism by imposing $f$ to be an isomorphism instead of  just a diffeomorphism. Note that this apparent restriction does not affect the estimation of the SO assisted electron coupling with flexural phonons, since we are introducing extrinsic curvature and setting the Gaussian curvature to zero.

\begin{figure}
\includegraphics[width=1\columnwidth]{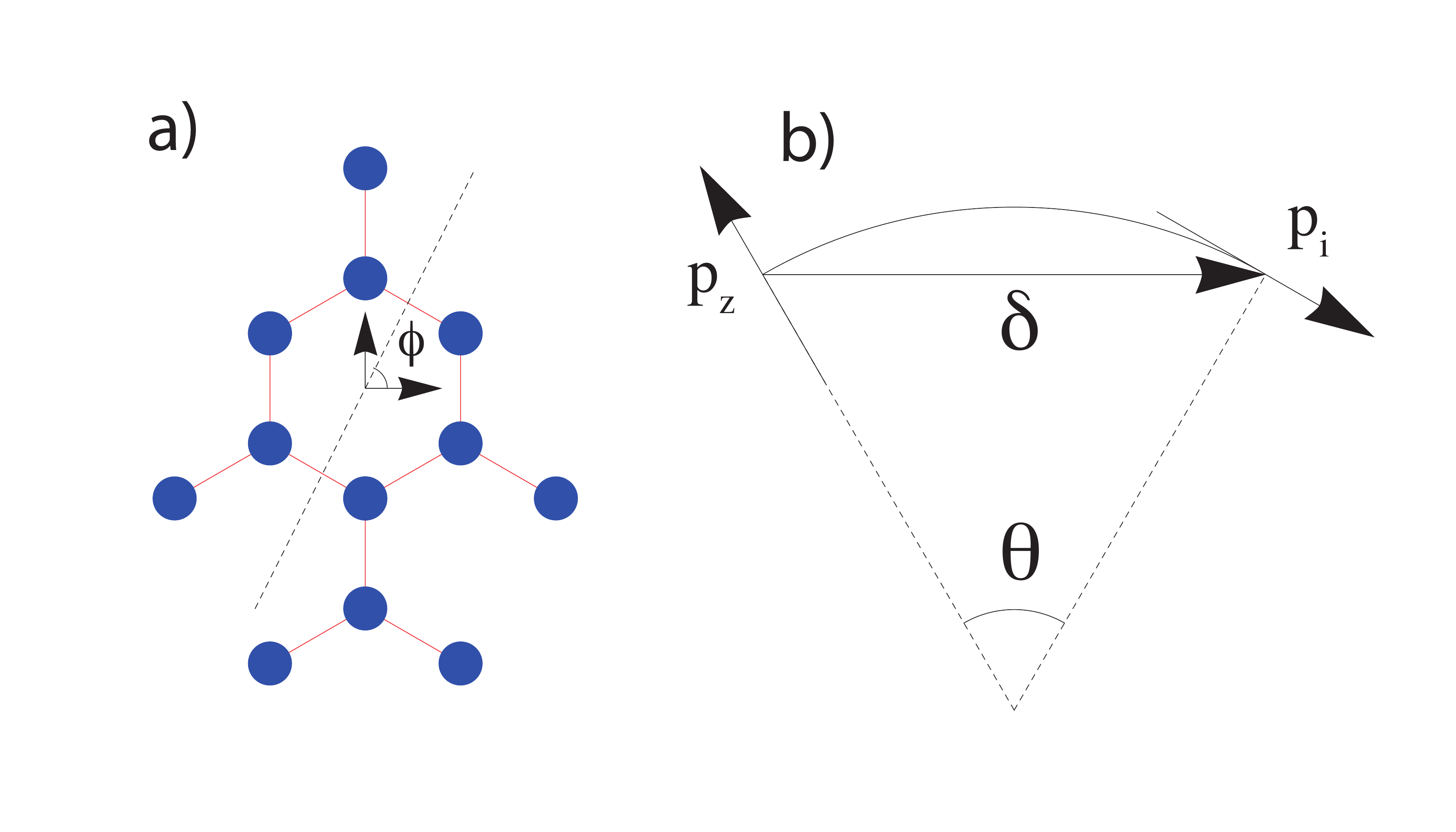}
\caption{\label{fig:cylinder}a) Definition of the angle $\phi$. b) Sketch for the calculation of the new hoppings between $p_z$ and $p_i$ orbitals.}
\end{figure}

We know how to choose the orbital basis and then calculate the two-center matrix elements, at least locally, using the Koster-Slater parametrization. For simplicity, we consider a curved graphene surface with a constant curvature along a given direction (a cylinder), so essentially the same problem as a carbon nanotube. Thus, we have two parameters, the radius of curvature $R$ and the angle $\phi$ between the direction of curvature and the $x$-axis (essentially the chiral angle in a nanotube), see Fig. \ref{fig:cylinder} a). The new hoppings between $\pi$ and $\sigma$ orbitals can be calculated following the prescription of Tab. \ref{tab:matrix_elements}. These  are function of the angle $\theta$ defined in Fig. \ref{fig:cylinder} b). Assuming that $R\gg a$, we have to the leading term in $a/R$:
\begin{align}
\theta\approx\frac{\left|\delta_x\cos(\phi)+\delta_y\sin(\phi)\right|}{R}
\end{align}
where $\delta_{x,y}$ are the components of the vector $\vec{\delta}$ which connects nearest neighbors. After a straightforward calculation, the block Hamiltonian that mixes $\pi$ and $\sigma$ states at $\mathbf{K}_{\pm}$ can be written as the matrix:
\begin{align}
\mathcal{H}_{\sigma\pi}=\frac{3a}{8R}\left(\begin{array}{cc}
0 & -V_{sp\sigma} e^{i\tau2\phi}\\
 -V_{sp\sigma} e^{-i\tau2\phi} & 0 \\
0 & i\tau\left(V_1+e^{-i\tau2\phi}V_2\right)\\
i\tau\left(V_1+e^{i\tau2\phi}V_2\right) & 0 \\
0 & V_1-e^{-i\tau2\phi}V_2\\
-V_1+e^{i\tau2\phi}V_2 & 0
\end{array}\right)
\end{align}
where $V_1=V_{pp\sigma}+V_{pp\pi}$, $V_2=\left(V_{pp\sigma}+3V_{pp\pi}\right)/2$, and $\tau=\pm 1$ labels the valley $\mathbf{K}_{\pm}$.

This expression is exact to the leading order in $a/R$ assuming a constant $R$ and $\phi$ along the graphene surface. Now we perform a local approximation, which is valid at long wavelengths: we assume that $R$ and $\phi$ depends slightly on the position.  Hence, they can be related with the second derivatives of the height profile, since the second fundamental form ($\mathcal{F}$) of a surface in the Monge's parametrization ($\mathbf{r}=(x,y,h(x,y))$) reads:
\begin{align}
\mathcal{F}=\frac{1}{\sqrt{1+\partial_i h\partial^i h}}\left(\begin{array}{cc}
\partial_x\partial_x h & \partial_x\partial_y h\\
\partial_x\partial_y h & \partial_y\partial_y h
\end{array}\right)
\end{align}
If we neglect quadratic terms on $h$, then $\mathcal{F}$ is just the tensor of second derivatives. At the same time, note that within the continuum description of graphene as a membrane, the height profile $h$ should be identified with the flexural acoustic phonon field at long wavelengths $u_{A_1}$. Then, the local $R$ and $\phi$ can be related with the second derivatives of $u_{A_1}$ as:
\begin{align}
\partial_x\partial_x u_{A_1}\approx -R^{-1}\cos^2\phi\nonumber\\
\partial_y\partial_y u_{A_1}\approx -R^{-1}\sin^2\phi\nonumber\\
\partial_x\partial_y u_{A_1}\approx -R^{-1}\sin\phi\cos\phi
\end{align}

By projecting out the $\sigma$ electronic states as Eq. \ref{eq:Heff} indicates, we arrive at the electron-phonon coupling of Eq. \ref{eq:e-A1} with:
\begin{align}
g_1=\frac{a\epsilon_s\Delta\left(V_{pp\sigma}+V_{pp\pi}\right)}{12V_{sp\sigma}^2}\approx3\,\,\mbox{meV}\cdot\AA\nonumber\\
g_2=\frac{aV_{pp\pi}\Delta}{2\left(V_{pp\sigma}-V_{pp\pi}\right)}\approx4\,\,\mbox{meV}\cdot\AA\nonumber\\
\end{align}
Importantly, within the present Koster-Slater approximation we have to extend the tight-binding calculation to second-nearest neighbors in order to obtain a non-zero $g_3$ coupling. In that case we obtain:
\begin{align}
g_3=\frac{a\Delta\left(3V_{pp\pi}+V_{pp\sigma}\right)\left(V^{(2)}_{pp\sigma}+V^{(2)}_{pp\pi}\right)}{8\left(V_{pp\sigma}-V_{pp\pi}\right)^2}
\end{align}
where $V^{(2)}_{pp\sigma}$ and $V^{(2)}_{pp\pi}$ are new second-nearest neighbors hopping parameters.

In order to estimate the coupling with optical phonons we have to consider the effect of a vertical displacement of one sublattice with respect to the other, similarly to the case of silicene.\citep{Falko_silicene} Processes which involve only one phonon give a vanishing contribution as expected from symmetry considerations. We can repeat the same scheme as before by considering virtual processes mediated by two phonons ($\propto \mathcal{H}_{\pi\sigma}^{flex}\mathcal{H}_{\sigma}^{-1}\mathcal{H}_{\sigma\pi}^{flex}$, note that $\mathcal{H}_{\sigma}$ also contains the SO interaction). We identify the Kane-Mele like coupling with flexural optical phonons at $\mathbf{\Gamma}$, whose strength reads:
\begin{align}
g_4=\frac{2\epsilon_s^2\Delta\left(V_{pp\pi}-V_{pp\sigma}\right)^2}{9a^2V_{sp\sigma}^4}\approx20\,\,\mbox{meV}\cdot\AA^{-2}
\end{align}

\subsection*{Phonons at $\mathbf{K}_{\pm}$}

We estimate now the coupling with phonons at the corners of the Brillouin zone. We simplify the tight-binding model in order to treat the problem analytically. We are going to consider the model described in Ref. \onlinecite{Thorpe_Weaire} and adapted in Ref. \onlinecite{Paco_phonons} in order to describe the acoustic phonon modes in graphite.
We consider two parameters: $V_{on}$, which is the on-site energy of the $\sigma$ orbitals, and $V_{hop}$, which is the hopping between $\sigma$ orbitals at nearest neighbors when the orbitals are maximally localized in the direction which links the two atoms, otherwise the hopping is taken to zero. These parameters can be estimated from the Slater-Koster parameters as:\begin{align}
\label{model_parameters1}
V_{on}=\frac{\epsilon_s-\epsilon_p}{3}\nonumber\\
V_{hop}=\frac{V_{ss\sigma}-2\sqrt{2}V_{sp\sigma}-2V_{pp\sigma}}{3}
\end{align}
This model was employed in Ref. \onlinecite{Huertas-Hernando_etal_prb} in order to estimate the SOC in graphene and carbon nanotubes. Within this model for flat graphene, the Kane-Mele coupling reads $\Delta_{I}^{flat}=3V_{on}\Delta^2/(4V_{hop}^2)$, in agreement with Ref. \onlinecite{Huertas-Hernando_etal_prb}. Note also that this estimation is numerically very close to the one of Ref. \onlinecite{Min_etal}.

We perform the calculation in the unit cell with 6 atoms. We can estimate the terms that mix $\pi$ and $\sigma$ states by considering the vertical displacement of one lattice respect to the other, as we mentioned before. By projecting out the $\sigma$ orbitals we arrive to a 6x6 effective Hamiltonian for the $\pi$ electronic states at the new $\mathbf{\Gamma}$, which can be seen as a matrix expressed in the monoelectronic basis $\left(|\mathbf{\Gamma}A_1\pi\rangle,|\mathbf{\Gamma}A_2\pi\rangle,|\mathbf{\Gamma}A_3\pi\rangle,
|\mathbf{\Gamma}B_1\pi\rangle,|\mathbf{\Gamma}B_2\pi\rangle,|\mathbf{\Gamma}B_3\pi\rangle\right)$. In order to identify the effective Hamiltonian in the low energy sector we have to express this matrix in the monoelectronic basis associated to the lattice with 2 atoms per unit cell, let's say $\left(|\mathbf{\Gamma}A\pi\rangle,|\mathbf{\Gamma}B\pi\rangle,|\mathbf{K}_+A\pi\rangle,
|\mathbf{K}_+B\pi\rangle,|\mathbf{K}_-A\pi\rangle,|\mathbf{K}_-B\pi\rangle\right)$. Both basis are related by the unitary transformation:\begin{align}
U=\frac{1}{\sqrt{3}}\left(\begin{array}{cccccc}
1 & 0 & 1 & 0 & 1 & 0\\
1 & 0 & e^{i\frac{2\pi}{3}} & 0 & e^{-i\frac{2\pi}{3}} & 0\\
1 & 0 & e^{-i\frac{2\pi}{3}} & 0 & e^{i\frac{2\pi}{3}} & 0\\
0 & 1 & 0 & 1 & 0 & 1\\
0 & 1 & 0 & e^{i\frac{2\pi}{3}} & 0 & e^{-i\frac{2\pi}{3}}\\
0 & 1 & 0 & e^{-i\frac{2\pi}{3}} & 0 & e^{i\frac{2\pi}{3}}\\
\end{array}\right)
\end{align}
By doing so, we identify the strength of the coupling with phonons at the corner of the Brillouin zone:
\begin{align}
g_5=\frac{\sqrt{3}V_{on}\Delta \left(\sqrt{2}V_{pp\sigma}-\sqrt{2}V_{pp\pi}+V_{sp\sigma}\right)}{2\sqrt{2}aV_{hop}^2}\approx4\,\,\mbox{meV}\cdot\AA^{-1}
\end{align}

\section{Enhancement of Kane-Mele mass}

One of the most interesting consequences of this analysis is the effect of the coupling of Eq. \ref{eq:e-B2} in the electronic spectrum. The contribution of flexural phonons to the Kane-Mele coupling can be written as:\begin{align}
\Delta_{ph}=g_4\left\langle(u_{B_2})^2\right\rangle
\label{eq:KM1}
\end{align}where the brackets express the thermal average over the entire Brillouin zone.

\begin{figure}
\includegraphics[width=1\columnwidth]{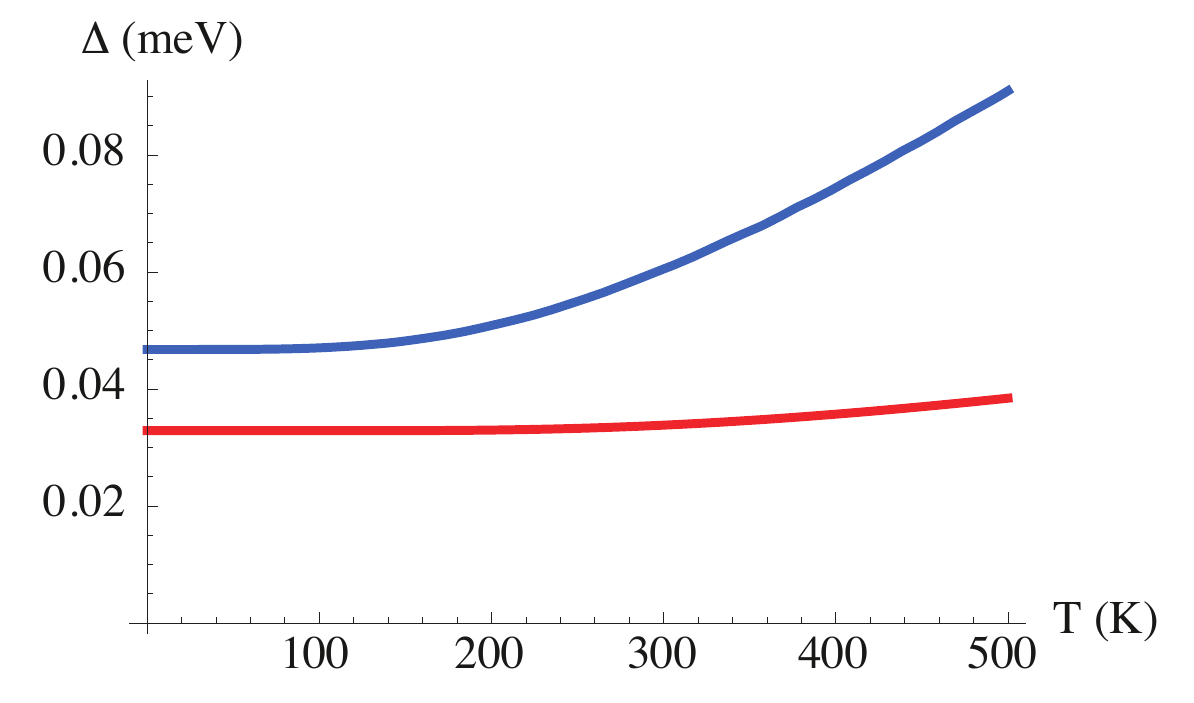}
\caption{\label{fig:figKM}Effective Kane-Mele mass induced by the coupling with flexural phonons. In red (lower curve) the estimation neglecting the acoustic branch and the dispersion of the optical one. In blue (upper curve) the calculation within the model described in Appendix B.}
\end{figure}

The flexural optical (ZO) mode strictly at $\mathbf{\Gamma}$ transforms according to $B_2$. A rough estimate consists on neglecting the contribution from the acoustic branch and the dispersion of the optical mode. Since $\hbar\omega^{ZO}_{\Gamma}\approx 110$ meV,\citep{Wirtz_Rubio} temperature plays no role, see the red curve in Fig. \ref{fig:figKM}. However, the zero-point motion contribution $\Delta_0=\frac{\hbar g_{4}}{2M\omega^{ZO}_{\Gamma}}\approx$ 0.03 meV is non-negligible (here $M$ is the mass of the carbon atom).

This is a very crude approximation, since the identification of $u_{B2}$ with the ZO mode is strictly true at the $\mathbf{\Gamma}$ point. The entire Brillouin zone contributes to the average, so away from $\mathbf{\Gamma}$ both flexural acoustic (ZA) and optical branches enter. We can use the symmetry-adapted basis $|A_1\rangle$ and $|B_2\rangle$ in order to describe the polarizations of the $\nu=$ ZA,ZO phonons: \begin{align}
|\nu\rangle=\eta_{A_1}^{\nu}\left(\mathbf{q}\right)|A_1\rangle+
\eta_{B_2}^{\nu}\left(\mathbf{q}\right)|B_2\rangle
\end{align}
Note that, because of time reversal symmetry, $\eta\left(\mathbf{q}\right)=\left[\eta\left(-\mathbf{q}\right)\right]^*$. Thus, we have:\begin{align}
\left\langle(u_{B_2})^2\right\rangle\approx\frac{1}{N}\sum_{\mathbf{q}\in BZ}\sum_{\nu}\left|\eta_{B_2}^{\nu}\left(\mathbf{q}\right)\right|^2\left\langle \left|u_{\mathbf{q}}^{\nu}\right|^2\right\rangle_T
\label{eq:KM2}
\end{align}where:\begin{align}
u_{\mathbf{q}}^{\nu}=\frac{1}{\sqrt{N}}\sum_{i=1}^N u^{\nu}\left(\mathbf{R}_i\right)e^{-i\mathbf{q}\cdot\mathbf{R}_i}
\end{align}
is the Fourier transform of the phonon displacement field in branch $\nu$, and the brackets denote thermal average.

We need a model in order to describe the deviations of the polarizations vectors and the frequencies of both branches in the entire Brillouin zone. In Appendix B we describe the simpler nearest-neighbor forces model that one can consider in order to describe the dynamics of flexural phonons. We compute the induced Kane-Mele mass from Eqs. \eqref{eq:KM1} and \eqref{eq:KM2} within this model. The results are shown in Fig. \ref{fig:figKM} (blue curve). Remarkably, the Kane-Mele gap induced by phonons $2\Delta_{ph}$ is of the order of 0.1 meV.

\section{Conclusions}
We have studied the possible SO-mediated electron interaction with flexural phonons allowed by the symmetries of the lattice and estimated the strength of these couplings from a tight-binding model. This analysis can be used in order to study the SO coupling in carbon nanotubes. 

We find that the quadratic coupling with the phonons at $\mathbf{\Gamma}$, particularly with the ZO branch, renormalizes the Kane-Mele mass in a remarkable way. From our theory, we predict an enhancement of two orders of magnitude in comparison to previous estimations for flat graphene, putting this gap close to the present experimental limits.\citep{Mayorov_etal} Note that the frequency of flexural modes depends on the coupling to the substrate, if any, and can be tuned by applied strains.\cite{AGM12} On general grounds, it can be expected that a compressive strain will lower the frequency of these modes, enhancing the spin-orbit coupling.

Our theory is also relevant for spin transport experiments, particularly in suspended samples,\citep{Guimaraes_etal} where charge transport is ultimately limited by flexural phonons.\citep{Castro_etal}

\section{Acknowledgements}
This work has been funded by the
MICINN, Spain, (FIS2008-00124, FIS2011-23713, CONSOLIDER CSD2007-00010), and ERC, grant 290846. AHCN acknowledges DOE grant DE-FG02-08ER46512, ONR grant MURI N00014-09-1-1063, and the NRF-CRP award "Novel 2D materials with tailored properties: beyond graphene" (R-144-000-295-281). H. O. acknowledges financial support through grant JAE-Pre (CSIC, Spain). The authors acknowledge the hospitality of the KITP, Santa Barbara. This work is partially supported by the National Science Foundation under Grant No. NSF PHY11-25915.
\appendix

\section{Choice of the spinor basis}

With this appendix we try to avoid possible confusions due to notation and choice of spinor basis in the main text, in particular when compared to Ref. \onlinecite{Kane_Mele}.

When discussing the symmetries of the graphene lattice from the perspective of group theory it is common to introduce electronic operators without specifying their explicit expression in a certain basis, only the algebraic rules which determine how they transform under the symmetry operations of the enlarged point group $C_{6v}''$. For instance, the two $4\times4$ matrices which transform according to the vector irreducible representation $E_1$ are denoted by $\Sigma_x$, $\Sigma_y$. The matrix $\Sigma_z$ is defined in order to complete the Pauli-matrix algebra, and it transforms according to $A_2$. On the other hand, the matrices which transform according to the vector irreducible representation $E_2'$ are denoted by $\Lambda_x$, $\Lambda_y$, and as before $\Lambda_z$ is defined in order to complete the Pauli-matrix algebra and transforms according to $B_1$.

\begin{center}
\begin{table}
\begin{tabular}{|c|c|c|}
\hline
Irrep&$z\rightarrow-z$ symmetric&$z\rightarrow-z$ asymmetric\\
\hline
$A_1$&$\sigma_z\otimes\tau_z\otimes s_z$ &$\sigma_x\otimes\tau_z\otimes s_y-\sigma_y\otimes s_x$\\
\hline
$A_2$& &$\sigma_x\otimes\tau_z\otimes s_x+\sigma_y\otimes s_y$\\
\hline
$B_2$&$\tau_z\otimes s_z$&\\
\hline
$E_1$&$\left(\begin{array}{c}
-\sigma_y\otimes s_z \\
\sigma_x\otimes\tau_z\otimes s_z \end{array}\right)$& $\left(\begin{array}{c}
-\sigma_z\otimes\tau_z\otimes s_y \\
\sigma_z\otimes\tau_z\otimes s_x \end{array}\right)$\\
\hline
$E_2$& &$\left(\begin{array}{c}
\sigma_x\otimes\tau_z\otimes s_y+\sigma_y\otimes s_x \\
\sigma_x\otimes\tau_z\otimes s_x-\sigma_y\otimes s_y \end{array}\right)$\\ &&$\left(\begin{array}{c}
-\tau_z\otimes s_y \\
\tau_z\otimes s_x \end{array}\right)$\\
\hline
$E_1'$&$\left(\begin{array}{c}
-\sigma_y\otimes\tau_y\otimes s_z \\
\sigma_y\otimes\tau_x\otimes s_z \end{array}\right)$&\\
\hline
$G'$& & $\left(\begin{array}{c}
-\sigma_y\otimes\tau_y\otimes s_x\\
-\sigma_y\otimes\tau_y\otimes s_y \\
\sigma_y\otimes\tau_x\otimes s_y \\
\sigma_y\otimes\tau_x\otimes s_x\end{array}\right)$\\
\hline
\end{tabular}
\caption{Classification of the possible SO coupling terms according to how they transform under the symmetry operations of the lattice and reflection in the graphene plane.}
\label{tab:so_appendix}
\end{table}
\end{center}

A more physical construction would be the following. As we mentioned in the text, in order to write down the low-energy electronic Hamiltonian (neglecting the spin) we have to consider the 16 Hermitian operators acting in a 4-dimensional space. We could define two different sets of $2\times2$ Pauli matrices $\left\{\sigma_i\right\}$, $\left\{\tau_i\right\}$ associated to the physical sublattice and valley degrees of freedom respectively, so the set $\left\{\mathcal{I},\sigma_i,\tau_i,\sigma_i\otimes\tau_j\right\}$ provides a representation of the algebra of generators of $U(4)$. These operators act in a space of 4-component Bloch functions $\Psi=(\psi_{A,\mathbf{K}_+},\psi_{B,\mathbf{K}_+},\psi_{A,\mathbf{K}_-,\uparrow},\psi_{B,\mathbf{K}_-})^T$. Then, we introduce the Pauli matrices associated to spin, so we have to double the space of Bloch functions in order to contain the two spin projections. The possible SO couplings in this basis are summarized in Tab. \ref{tab:so_appendix}. In the first row we can recognize the Kane-Mele and Rashba-like couplings as discussed in Ref. \onlinecite{Kane_Mele}. Note that in this basis the time reversal operation is implemented by the anti-unitary operator $\mathcal{T}=i\tau_x\otimes s_y\mathcal{K}$, where $\mathcal{K}$ denotes complex conjugation.

In the main text we employ a different basis, where the order of the projection of the Bloch functions at each sublattice in different valleys is inverted, and also a minus sign is introduced: $\Psi=(\psi_{A,\mathbf{K}_+},\psi_{B,\mathbf{K}_+},\psi_{B,\mathbf{K}_-},-\psi_{A,\mathbf{K}_-})^T$. This basis is very convenient because the notation is simplified. In this basis the operators $\Sigma_i$, $\Lambda_i$ are related with the matrices acting in subalattice and valley indices as:\begin{align}
\Sigma_i=\sigma_i\otimes \mathcal{I}\nonumber\\
\Lambda_i=\mathcal{I}\otimes\tau_i
\end{align}The notation is simplified essentially because in the basis the time reversal operation is implemented by $\mathcal{T}=i\Sigma_y\otimes\Lambda_y\otimes s_y\mathcal{K}$, in such a way that the three sets of matrices are odd under the action of $\mathcal{T}$:\begin{align}
\Sigma_i\xrightarrow{\;\mathcal{T}\;}\Sigma_y\Sigma_i^{*}\Sigma_y=-\Sigma_i\nonumber\\
\Lambda_i\xrightarrow{\;\mathcal{T}\;}\Lambda_y\Lambda_i^{*}\Lambda_y=-\Lambda_i\nonumber\\
s_i\xrightarrow{\;\;\;\mathcal{T}\;\;\;} s_ys_i^{*}s_y=-s_i
\end{align}and then the possible SO terms are constructed from products of a spin matrix $s_i$ with $\Sigma_i$ or $\Lambda_i$.

\section{Model for flexural phonons}

We consider the simpler nearest-neighbor forces model where the elastic energy of the lattice can be written as:
\begin{align}
E=\frac{\alpha}{a^2}\sum_{i}\left[\left(h_{Ai}-\frac{1}{3}\sum_{\langle ij\rangle}h_{Bj}\right)^2+\left(h_{Bi}-\frac{1}{3}\sum_{\langle ij\rangle}h_{Aj}\right)^2\right]
\end{align}
Here $a$ is the carbon-carbon distance and $\alpha$ is a constant with units of energy which can be related with the bending rigidity ($\kappa$) of graphene in a continuum description,\citep{Landau_elasticity} as we are going to see next. This model leads to the dynamical matrix:
\begin{align}
\mathcal{D}\left(\mathbf{q}\right)=\frac{2\alpha}{3a^2}\left(\begin{array}{cc}
3+\frac{|f\left(\mathbf{q}\right)|^2}{3} & -2f\left(\mathbf{q}\right) \\
-2f\left(\mathbf{q}\right)^* & 3+\frac{|f\left(\mathbf{q}\right)|^2}{3}
\end{array}\right)
\end{align}
where $f\left(\mathbf{q}\right)=\sum_{\alpha}e^{i\mathbf{q}\cdot\vec{\delta}_{\alpha}}$, and the sum is extended to nearest-neighbors. The frequencies read:
\begin{align}
\omega_{\pm}=\sqrt{\frac{2\alpha}{3Ma^2}\left(3+\frac{|f\left(\mathbf{q}\right)|^2}{3}\pm2|f\left(\mathbf{q}
\right)|\right)}
\label{eq:wpm}
\end{align}
with polarization vectors $|\pm\rangle=\frac{1}{\sqrt{2}}\left(\frac{f\left(\mathbf{q}\right)}
{\left|f\left(\mathbf{q}\right)\right|},\mp1\right)^T$. Here $M$ is the mass of the carbon atom. The two branches of Eq. \eqref{eq:wpm} are plotted in Fig. \ref{fig:z_dispersion}. The upper branch $\omega_{+}$ must be identified with the optical one, whereas $\omega_-$ corresponds to the acoustic one.

\begin{figure}
\begin{centering}
\includegraphics[width=1\columnwidth]{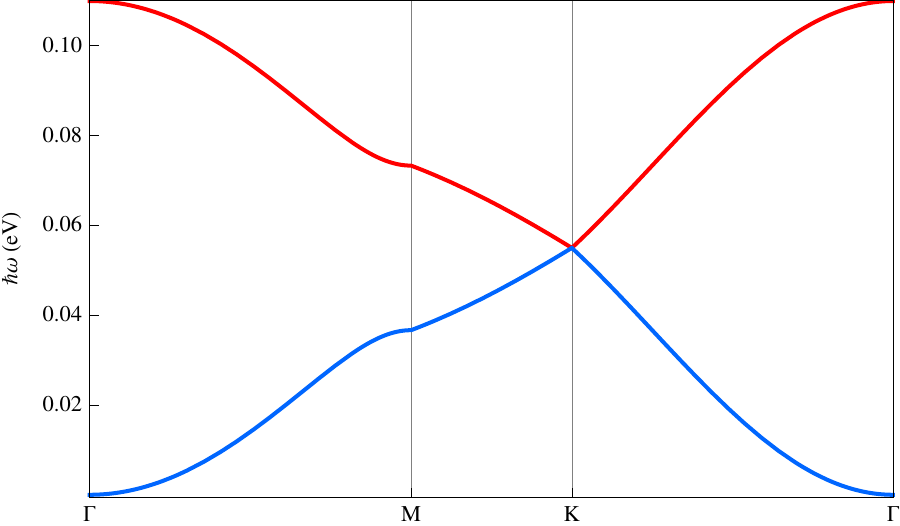}
\par\end{centering}
\caption{\label{fig:z_dispersion}Dispersion of flexural phonons computed within the nearest-neighbor force model described in the text with $\alpha=8.5$ eV. In red (upper curve) the dispersion for the optical branch, in blue (lower curve) the acoustic branch.}
\end{figure}

The model reproduces very well the dispersion of flexural phonons.\citep{Wirtz_Rubio} At $\mathbf{q}\sim\mathbf{\Gamma}$ we have:
\begin{align}
\omega_+\equiv\omega_{\mathbf{q}}^{ZO}\approx\sqrt{\frac{8\alpha}{Ma^2}}-\sqrt{\frac{\alpha a^2}{8M}}q^2\nonumber\\
\omega_-\equiv\omega_{\mathbf{q}}^{ZA}\approx\sqrt{\frac{\alpha}{8Ma^{-2}}}q^2
\end{align}
Note that the dispersion relation of ZA phonons is quadratic, as expected from symmetry considerations. On the other hand, both branches are degenerate at $\mathbf{K}_{\pm}$ ($\omega_{\mathbf{K}_{\pm}}^{ZA}=\omega_{\mathbf{K}_{\pm}}^{ZO}
=\omega_{\mathbf{\Gamma}}^{ZO}/2$), as it is also expected from symmetry arguments. When we compare this model with the theory of elasticity\citep{Landau_elasticity} we deduce the relation:\begin{align}
\alpha=6\sqrt{3}\kappa
\label{eq:relation}
\end{align}

We set the value of $\alpha$ from the frequency of the flexural optical phonon at $\mathbf{\Gamma}$, $\hbar\omega_{\mathbf{\Gamma}}^{ZO}\approx110$ meV.\citep{Wirtz_Rubio} We obtain $\alpha=8.5$ eV. By using the relation of Eq. \eqref{eq:relation} we obtain $\kappa\approx 0.8$ eV, which is a very reasonable value for the bending rigidity of graphene. This agreement confirms the reliability of the model. We compute the induced Kane-Mele mass from Eqs. \eqref{eq:KM1} and \eqref{eq:KM2} within this model. Note that:
\begin{align}
\left|\eta_{B_2}^{ZO}\left(\mathbf{q}\right)\right|^2=\frac{1}{2}\left(1+\frac{\Re f\left(\mathbf{q}
\right)}{\left|f\left(\mathbf{q}
\right)\right|}\right)
\nonumber\\
\left|\eta_{B_2}^{ZA}\left(\mathbf{q}\right)\right|^2=\frac{1}{2}\left(1-\frac{\Re f\left(\mathbf{q}
\right)}{\left|f\left(\mathbf{q}
\right)\right|}\right)
\end{align}
where $\Re f$ denotes the real part of $f$. The results are shown in Fig. \ref{fig:figKM} (blue curve).

\bibliography{Bibliography/ti,Bibliography/graphene_so,Bibliography/graphene,Bibliography/graphene_spintronics,Bibliography/majoranas,Bibliography/books,Bibliography/techniques}
\end{document}